\documentclass[twocolumn, showpacs,preprintnumbers,amsmath,amssymb]{revtex4}

 \usepackage{graphicx}
\usepackage{verbatim}

\begin{document}

\addtolength{\textheight}{1.2cm}
\addtolength{\topmargin}{-0.5cm}

\newcommand{\etal} {{\it et al.}}

\title{Detection through exchange energy of multipartite entanglement in spin rings}

\author{I. Siloi$^{1,2}$ and F. Troiani$^2$}

\affiliation{$^{1}$ Dipartimento di Fisica, Universit\`a di Modena e Reggio Emilia, 
Italy}
\affiliation{$^{2}$ S3 Istituto Nanoscienze-CNR, Modena, Italy}

\date{\today}

\begin{abstract}

We investigate multipartite entanglement in rings of arbitrary spins with antiferromagnetic interactions between nearest neighbors. 
In particular, we show that the non-degenerate ground state of rings formed by an even number ($N$) of spins is $N$-partite entangled, and exchange energy can thus be used as a multipartite-entanglement witness.
We develop a general approach to compute the energy minima corresponding to biseparable states, and provide numerical results for a representative set of systems.  
Despite its global character, exchange energy also allows a spin-selective characterization of entanglement. In particular, in the presence of a magnetic defect, one can derive separability criteria for each individual spin, and use exchange energy for detecting entanglement between this and all the other spins.

\end{abstract}


\maketitle

Entanglement is a peculiar feature of composite quantum systems and has been recognized as a key resource in many fields of quantum technology \citep{nielsen_chuang01,Hor+09rmp}. 
In particular, the generation and detection of multipartite entanglement (ME) has been recently obtained in diverse physical systems \citep{guhne2009entanglement,kiesel2007experimental,wieczorek2009experimental,lu2007experimental,monz2011qubit,hammerer2010rev,neumann2008multipartite}. 
Besides their fundamental interest, multipartite entangled states have important applications in quantum computation \citep{Raussendorf,briegel2009measurement} and in quantum metrology, where they allow one to reach high sensitivities, beyond the standard quantum limit \citep{giovannetti2004quantum,sorensen2001many,hyllus2012fisher,toth2012multipartite}. 
More generally, ME can also be found in the ground and thermal states of prototypical many-body systems, such as spin systems with isotropic, $XY$ and noncollinear Ising interactions \citep{Amc+08rmp,wan02pra,bruss2005multipartite,guhne_toth06pra,troiani2011entanglement}.

A practical tool for detecting (multipartite) entanglement is represented by entanglement witnesses \citep{terhal2000bell,lewenstein2000optimization}. Such operators are defined such that their expectation value can exceed a given threshold only in the presence of some specific form of entanglement \citep{acin2001classification,bourennane2004experimental,huber2010detection,duan2011entanglement,sperling2013multipartite}. 
As the knowledge of the state is not required in this approach, entanglement witnesses are especially convenient with condensed matter systems, where such knowledge is in general unattainable, due to the large number of degrees of freedom \citep{krammer2009multipartite,cramer2011measuring}.
In spin systems, routinely measured thermodynamic quantities - such as the magnetic susceptibility \cite{brukner2006crucial}, inelastic neutron scattering \cite{wiesniak2005magnetic}, and exchange energy \cite{brukner2004macroscopic,wang2002quantum} - correspond to witnesses for spin-pair entanglement. Thus, the experimental detection of entanglement is allowed even in systems where local constituents cannot be locally addressed. Along the same lines, spin squeezing inequalities allow the detection of multipartite entanglement in the vicinity of specific states, through the measurement of collective spin operators \citep{sorensen2001entanglement,korbicz2005spin,toth2007optimal,vitagliano2011spin}.

Exchange energy also represents a witness for block \cite{Troiani13} and multipartite entanglement \citep{dowling+04pra,guhne+05NJOP}. In particular, it was shown that the ground state of an even-$N$ spin chain, with nearest neighboring Heisenberg interactions, cannot be written in any biseparable form, and is thus $N$-partite entangled \cite{troiani2012energy}. 
This general property allows the use of exchange energy for the detection of $N$-partite entanglement. Its presence can in fact be inferred from any expectation value of the exchange Hamiltonian that falls in the range between the ground state energy ($E_{0}$) and the lowest value achievable by a biseparable state ($E_{bs}$). In the present paper, we show that such approach can also be applied to spin rings. Besides their interest as prototypical models of highly entangled quantum systems \citep{wooters2001entangled,arnsen+01prl}, 
spin rings have found a large number of physical implementations in molecular magnetism \cite{Gatteschi}. Amongst various rings of antiferromagnetically-coupled transition-metal ions, a particular attention was devoted in the last years to the more restricted class of Cr-based systems \cite{Winpenny07}. Here, the ability of tailoring the physical properties at the synthetic level also results in the possibility of engineering the entanglement features \cite{Lorusso12,siloi2012towards,candini+10prl}. 

In order to enable the detection of multipartite entanglement in molecular spin clusters, we develop a flexible approach, which applies to systems of arbitrary spin. This allows us to compute the minimum energy attainable by a biseparable state ($E_{bs}$) in a variety of spin rings. In particular, for any given partition of the ring in two complementary subsystems (hereafter labeled $A$ and $B$), $E_{bs}$ is found as a self-consistent solution of two spin-chain Hamiltonians, one for each of the subsystems. The coupling between the uncorrelated spins at the boundaries of $A$ and $B$ is described in terms of an effective, local magnetic field. Even though the parameter space is significantly larger than in the case of spin chains, we show that it can be partially reduced on the basis of intuitive arguments, whose validity is numerically verified in a number of test cases. 
Finally, we show that the overall exchange energy can be used in systems formed by inequivalent spins (such as spin chains or rings with magnetic defects) in order to detect the presence of entanglement between each spin in the system and all the others. Therefore, quite remarkably, a collective observable enables a spin-selective investigation of entanglement in the system.\\
The paper is organized as follows. Section \ref{sec1} is dedicated to the use of exchange energy as a multipartite entanglement witness. We first demonstrate that the ground state of an even-$N$ spin ring is $N$-partite entangled (Sec. \ref{sec1a}). We then calculate the energy minima for biseparable states, in the specific case of $N=8$ and for different spin lengths $s=1/2,1,3/2$ (Sec. \ref{sec1b}). In Sec. \ref{sec2} we consider inhomogeneous systems, where exchange energy can be used as a probe of local properties. We specifically refer to a class of heterometallic molecular nanomagnets, namely the Cr-based rings with different chemical substitutions.

\section{Energy as a witness of multipartite entanglement}
\label{sec1}

In the present Section, we demonstrate that the ground state $ | \Psi_0 \rangle $ of an $N$-spin ring (for even $N$), with exchange coupling between nearest neighbors, is $N$-partite entangled. In other words, such ground state cannot be written in any biseparable form $ | \Psi_0 \rangle = | \Psi_A \rangle \otimes | \Psi_B \rangle $, with $A$ and $B$ any two complementary subsystems. 
This property, along with the fact that the ground state is non-degenerate, always allows the detection of multipartite entanglement in these systems. In particular, the presence of such entanglement can be deduced from expectation values of the exchange energy lower than given thresholds. A general approach for numerically calculating such thresholds is derived in the second part of the Section.

\subsection{Multipartite entanglement in the ground state of spin rings}
\label{sec1a}
{\it Theorem. ---} The ground state $|\Psi_0\rangle$ of the spin Hamiltonian 
$ H = \sum_{i=1}^{N} {\bf s}_i \cdot {\bf s}_{i+1}$, with even $N$, cannot be written 
in any biseparable form $|\Psi_{bs}\rangle = |\Psi_A\rangle \otimes |\Psi_B\rangle$, 
and is thus $N$-partite entangled.

{\it Proof. ---} 
The first two steps of the demonstration, which we recall hereafter for completeness, coincide with the ones that apply to the spin chains \cite{troiani2012energy}. According to Marshall's theorem, $|\Psi_0\rangle$ is a non-degenerate $S=0$ state \cite{auerbach}.

A biseparable state $ |\Psi_{bs}\rangle $ can only be in a singlet state ($S=0$) if the same applies to each of the subsystems $A$ and $B$ ($S_A=S_B=0$). In fact, if one writes $ |\Psi_\chi\rangle $ ($ \chi = A,B $) as a linear superposition of eigenstates of ${\bf S}_\chi^2$, $|\Psi_\chi\rangle = \sum_{S_\chi} C_{S_\chi}^\chi | \phi^\chi_{S_\chi} \rangle $, then one can show that:
$ \langle {\bf S}^2\rangle 
\ge
\sum_{S_A,S_B} |C_{S_A}^A C^B_{S_B}|^2 [ (S_A - S_B)^2 + S_A + S_B ] 
\ge
\sum_{S_A,S_B} |C_{S_A}^A C^B_{S_B}|^2 ( S_A + S_B )$,
being
$ \langle \phi^A_{S_A} | {\bf S}_A | \phi^A_{S_A} \rangle \cdot 
  \langle \phi^B_{S_B} | {\bf S}_B | \phi^B_{S_B} \rangle \ge 
- S_A S_B $. 
As a consequence, $ \langle {\bf S}^2\rangle = 0$ implies 
$C^A_{S_A}=\delta_{S_A,0}$ and $C^B_{S_B}=\delta_{S_B,0}$.

As a final step, we prove that the state $ |\Psi_A\rangle \otimes |\Psi_B\rangle $, with 
$S_A=S_B=0$, cannot be the ground state of $H$. The relevant case is that where $A$ consists of the $N_A$ consecutive spins $s_1, s_2, \dots , s_{N_A}$, and $B$ of the and $N_B=N-N_A$ spins  $s_{N_A+1}, s_{N_A+2}, \dots , s_N$.
We write the spin Hamiltonian of the ring as the sum of three terms: 
$ H = H_A + H_B + H_{AB} $,
where 
$ H_A = \sum_{i=    1}^{N_A-1} {\bf s}_i \cdot {\bf s}_{i+1} $,
$ H_B = \sum_{i=N_A+1}^{N  -1} {\bf s}_i \cdot {\bf s}_{i+1} $,
and 
$ H_{AB} = {\bf s}_{N_A} \cdot {\bf s}_{N_A+1} 
         + {\bf s}_{N  } \cdot {\bf s}_{    1} $.
In the partial spin sum basis \cite{tsukerblat}, the state of $A$ reads:
\[
| \Psi_A \rangle = \sum_{\alpha,x_A} D_{\alpha,x_A,y_A} | \alpha , x_A , y_A=s, S_{A} =0, M_{A}=0 \rangle , 
\]
where $ \alpha $ denotes the quantum numbers 
$ S_1, \dots S_{N_A-3} $ corresponding to the partial spin sums
${\bf S}_k \equiv \sum_{i=2}^{k+1} {\bf s}_k$,
whereas $x_A = S_{N_A-2}$ and $y_A = S_{N_A-1}$ 
correspond to the partial spin sums
$ {\bf S}_{N_A-2} \equiv {\bf S}_{N_A-3} + {\bf s}_{N_A-2} $ 
and
$ {\bf S}_{N_A-1} \equiv {\bf S}_{N_A-2} + {\bf s}_{1    } $; 
the last spin to be summed in this coupling scheme is thus 
${\bf S}_{N_A} = {\bf S}_{A} - {\bf S}_{N_A-1}$.
The equation $S_A=0$ implies that $S_{N_A-1} = s$.
The same applies to the subsystem $B$, where the the spins at the boundary are summed last, and the state is expressed as:
\[
| \Psi_B \rangle = \sum_{\beta,x_B} D_{\beta,x_B,y_B} | \beta , x_B , y_B=s, S_{B} =0, M_{B}=0 \rangle , 
\]
where $ \beta $ denotes the quantum numbers 
$ S_1, \dots S_{N_B-3} $ corresponding to the partial spin sums
${\bf S}_k \equiv \sum_{i=N_A+2}^{N_A+1+k} {\bf s}_k$,
whereas $x_B = S_{N_B-2}$ and $y_B = S_{N_B-1}$ 
correspond to the partial spin sums
$ {\bf S}_{N_B-2} \equiv {\bf S}_{N_B-3} + {\bf s}_{N-2} $ 
and
$ {\bf S}_{N_B-1} \equiv {\bf S}_{N_B-2} + {\bf s}_{N  } $;
the last spin to be summed in this coupling scheme is thus 
${\bf S}_{N_A+1} = {\bf S}_{B} - {\bf S}_{N_B-1}$.

In order to demonstrate that $ |\Psi_A\rangle \otimes |\Psi_B\rangle $ cannot be an eigenstate of $H$, we prove that $ H |\Psi_{bs}\rangle $ has a finite component $ |\Psi_{bs}^\perp \rangle $ outside of the $S_A=S_B=0$ subspace, and thus orthogonal to $ |\Psi_{bs}\rangle $ itself.
We first show that 
$ | \Psi_{bs}^\perp \rangle \equiv | \Psi_{bs}^{\perp,1} \rangle +
                                   | \Psi_{bs}^{\perp,2} \rangle  
                            \equiv  (s_{z,N_A} s_{z,N_A+1}+
                                     s_{z,N  } s_{z,    1}) | \Psi_{bs} \rangle $ 
is finite and belongs to the subspace $S_{A/B}=1$ and $M_{A/B}=0$. We compute separately the two contributions 
$ | \Psi_{bs}^{\perp,1} \rangle $ and $ | \Psi_{bs}^{\perp,2} \rangle $, 
starting from the first one.
The operator $ s_{z,N_A} $ commutes with all ${\bf S}_k^2$ with $ k \le N_A-1$.
The matrix elements of the $N_A-$th spin can thus be reduced to those between the states of two spins $s$: 
$ s_{z,N_A} | \Psi_A \rangle = -\eta_s \sum_{\alpha,x_A} D_{\alpha,x_A,s} | \alpha , x_A , y_A=s, S_{A} =1, M_{A}=0 \rangle $, where 
$ \eta_s = [ ( \sum_{m=-s}^s m^2 ) / (2s+1)]^{1/2} > 0 $.
The same procedure is applied to $B$, resulting in:
$ s_{z,N_A+1} | \Psi_B \rangle = - \eta_s
\sum_{\beta,x_B} D_{\beta,x_B,s} | \beta , x_B , y_B=s, S_{B} =1, M_{B}=0 \rangle $. 
Therefore, 
\begin{align}\label{eq01}
| \Psi_{bs}^{\perp,1} \rangle \!
= \! \eta_s^2 \!\!\!\!\!
& \sum_{\alpha,\beta,x_A,x_B} \!\!\!\!\! D_{\alpha,x_A,s} D_{\beta ,x_B,s} \nonumber\\
&| \alpha, x_A , y_A\!=\!s, 1, 0 \rangle \otimes 
| \beta , x_B , y_B\!=\!s, 1, 0 \rangle .
\end{align}

The evaluation of $ | \Psi_{bs}^{\perp,2} \rangle $ is less straightforward. In order to simplify the calculation, we switch to a basis where the two spins at the boundary of each subsystem are summed in a reversed order with respect to the above coupling scheme: 
$ | \Psi_A \rangle = \sum_{\alpha,x_A} (-1)^{2s+x_A} D_{\alpha,x_A,s} | \alpha , x_A , y_A'=s, S_{A} =0, M_{A}=0 \rangle $. Here, $ \alpha $ and $x_A$ have the same meaning as above, whereas
$y_A' = S_{N_A-1}'$ corresponds to the partial spin sum
$ {\bf S}_{N_A-1}' \equiv {\bf S}_{N_A-2} + {\bf s}_{N_A} $; 
the last spin of $A$ to be summed in this coupling scheme is thus $s_1$.
The sign factor in the expression of $ | \Psi_A \rangle $ comes from the scalar product:
\[
\langle x_A , y_A=s , S_A = 0 | x_A , y_A' , S_A = 0 \rangle = \delta_{y_A',s} (-1)^{2s+x_A} .
\]
The same applies to $B$:
$ | \Psi_B \rangle = \sum_{\beta,x_B} (-1)^{2s+x_B} D_{\beta,x_B,s} | \beta , x_B , y_B'=s, S_{B} =0, M_{B}=0 \rangle $, where $ \beta $ and $x_B$ have the same meaning as above, whereas
$y_B' = S_{N_B-1}'$ 
corresponds to the partial spin sum
$ {\bf S}_{N_B-1}' \equiv {\bf S}_{N_B-2} + {\bf s}_{N_A+1} $;
the last spin of $B$ to be summed in this coupling scheme is thus $s_{N}$.
In order to verify that $ | \Psi_{bs}^{\perp} \rangle $ has a finite norm, we express 
$ | \Psi_{bs}^{\perp,2} \rangle $ in the same basis as $ | \Psi_{bs}^{\perp,1} \rangle $ (Eq. \ref{eq01}). In particular, we need to estimate its component along the $y_A=y_B=s$ subspace, to which $ | \Psi_{bs}^{\perp,1} \rangle $ belongs. 
The scalar product that allows to switch from one basis to the other within the relevant subspace is:
\begin{align*} 
\langle x_A , y_A=s , & S_A = 1 , M_A | x_A , y_A'=s , S_A = 1 , M_A \rangle = \\
& (-1)^{2s+1+x_A} \frac{2s(s+1)-x_A(x_A+1)}{2s(s+1)} .
\end{align*}
This results in the following expression:
\begin{align}  
| \Psi_{bs}^{\perp} & \rangle_{y_A=y_B=s}  
= \eta_s^2 
\sum_{\alpha,\beta,x_A,x_B} D_{\alpha,x_A,s} D_{\beta ,x_B,s} f(x_A,x_B) \nonumber\\ 
& 
| \alpha, x_A , y_A=s, 1, 0 \rangle \otimes 
| \beta , x_B , y_B=s, 1, 0 \rangle .
\end{align}
The factor $f$ is given by the expression:
\begin{equation}\label{eq02}
f(x_A,x_B) = 1 + \prod_{x=x_A,x_B}(-1)^{x} 
\left[ 1 - \frac{x(x+1)}{2s(s+1)} \right] ,
\end{equation}
where the first, constant term comes from $| \Psi_{bs}^{\perp,1} \rangle $ (see Eq. \ref{eq01}). 
In order to show that $| \Psi_{bs}^{\perp} \rangle$ has a finite norm, it suffices to show that this is the case for its projection in the $ y_A=y_B=s $ subspace.
We note that at least one of the products $ D_{\alpha,x_A,s} D_{\beta ,x_B,s} $ has to be finite, otherwise the norm of $ | \Psi_{bs} \rangle $ would vanish.
Besides, one can show that for all the relevant values of $x_A$ and $x_B$, namely $ 0 \le x_A , x_B \le 2s $, the function $f(x_A,x_B)$ is non-zero, i.e. the second term in Eq. \ref{eq02} differs from $-1$. 
In fact, for $x_A=x_B=0$ the terms in square brackets reduce to 1, and $f=2$. For all the other values of $(x_A,x_B)$, the modulus of the second term in $f$ is smaller than 1, and thus $f \neq 0$. 
This implies that $| \Psi_{bs}^{\perp} \rangle_{y_A=y_B=s}$, and thus 
$ | \Psi_{bs}^{\perp} \rangle $, has a finite norm.

The rest of the demonstration proceeds as in the case of the spin chain.
One can show that $ | \Psi_{bs}^\perp \rangle $ coincides with the component of 
$H | \Psi_{bs} \rangle $ with $S_A=S_B=1$ and $M_A=M_B=0$.
In fact, 
$ (H_A+H_B) | \Psi_{bs} \rangle $
belongs to the $S_A=S_B=0$ subspace,
being
$ [H_\chi,{\bf S}^2_{\chi'}] = 0 $ for $\chi, \chi' = A,B$.
The states
$ (s_{+,N_A} s_{-,N_A+1} + s_{+,1} s_{-,N}) | \Psi_{bs} \rangle $
and
$ (s_{-,N_A} s_{+,N_A+1} + s_{-,1} s_{+,N}) | \Psi_{bs} \rangle $
belong instead to the subspaces 
$ M_A = - M_B = + 1 $
and
$ M_A = - M_B = - 1 $,
respectively.
As a consequence, $ H| \Psi_{bs} \rangle $ has a finite component $ | \Psi^\perp_{bs}\rangle $, and thus cannot be an eigenstate of $H$.
$ \blacksquare $

\subsection{Energy minima for biseparable states}
\label{sec1b}
In systems formed by $N$ exchange-coupled spins, 
$ H = \sum_{i=1}^{N-1} {\bf s}_i \cdot {\bf s}_{i+1} $, 
energy allows the detection of genuine multipartite entanglement \cite{troiani2012energy}. 
This can be done by deriving the energy minimum $E_{bs}$ for biseparable states, 
$ | \Psi_{bs} \rangle = | \Psi_A \rangle \otimes | \Psi_B \rangle $,
where $A$ and $B$ are two subsystems into which the chain is partitioned. 
Here, any state $ | \Psi \rangle $ that violates the inequality: 
\begin{equation}\label{eq01}
\langle \Psi | H | \Psi \rangle \ge E_{bs}, 
\end{equation}
is $N$-partite entangled.
The first step in the derivation of the above minimum, is the calculation of the minimum $E_{bs} (N_A,N_B)$ corresponding to each given bipartition, where $N_A$ ($ N_B = N - N_A $) is the number of consecutive spins that form the subsystem $A$ ($B$). The lower bound in Eq. \ref{eq01} is the lowest such minima:
\begin{equation}
E_{bs} = \min_{N_A,N_B} E_{bs} (N_A,N_B) .
\end{equation}

For any given value of $N_A$, the Hamiltonian can be written as 
$ H = H_A + H_B + H_{AB} $, where:
\begin{eqnarray} 
H_A    & = & \sum_{i=1}^{N_A-1} {\bf s}_i \cdot {\bf s}_{i+1} , \ 
H_B      =   \sum_{i=N_A+1}^{N-1} {\bf s}_i \cdot {\bf s}_{i+1} \label{HABa}\\
H_{AB} & = & {\bf s}_{N_A} \cdot {\bf s}_{N_A+1} . \label{HABb}
\end{eqnarray}
The corresponding expectation value for a factorized state $ | \Psi_{bs} \rangle = | \Psi_A \rangle \otimes | \Psi_B \rangle $ reads:
\begin{eqnarray}
\langle \Psi_{bs} | H | \Psi_{bs} \rangle & = & \langle \Psi_A | H_A | \Psi_A \rangle + 
                      \langle \Psi_B | H_B | \Psi_B \rangle \nonumber\\
& + & 
\langle \Psi_A | {\bf s}_{N_A  } | \Psi_A \rangle
\cdot 
\langle \Psi_B | {\bf s}_{N_A+1} | \Psi_B \rangle .
\end{eqnarray} 
It can be shown that the minimum $E_{bs} (N_A,N_B)$ corresponds to the case where the expectation values of ${\bf s}_{N_A}$ and ${\bf s}_{N_A+1}$ are antiparallel to each other \cite{troiani2012energy}. This simplifies the search of the minimum as the (lowest) self-consistent solution of the two coupled eigenvalue problems, related to the Hamiltonians:
\begin{equation}\label{eq02}
\tilde{H}_A (z_B)\! =\! H_A\!+\!z_B s_{z,N_A}, \ \tilde{H}_B (z_A)\! =\! H_B\!+\!z_A s_{z,N_A+1}, 
\end{equation}
where $z_A \equiv \langle s_{z,N_A} \rangle $ and $z_B \equiv \langle s_{z,N_A+1} \rangle $ are the expectation values corresponding to the ground states of $\tilde{H}_A$ and $\tilde{H}_B$, respectively \cite{troiani2012energy}.

Hereafter, we apply the same procedure to the case of $N$ exchange-coupled spin rings, 
$ H = \sum_{i=1}^{N} {\bf s}_i \cdot {\bf s}_{i+1} $ (with ${\bf s}_{N+1} \equiv {\bf s}_1$). Unlike the case of spin chains, the subsystems $A$ and $B$ are coupled to each other through two spin pairs. Therefore, while Eq. \ref{HABa} applies to the present case as is, Eq. \ref{HABb} is replaced by:
\begin{equation}
H_{AB} = {\bf s}_{N_A} \cdot {\bf s}_{N_A+1} + {\bf s}_{N} \cdot {\bf s}_{1}.
\end{equation}
The two coupled Hamiltonians in Eq. \ref{eq02} are correspondingly replaced by: 
\begin{eqnarray}\label{selfcon}
\tilde{H}_A ({\bf z}_B,{\bf z}_B') & = & H_A + {\bf z}_B  \cdot {\bf s}_{N_A  } 
                      + {\bf z}_B' \cdot {\bf s}_{  1  } \nonumber\\
\tilde{H}_B ({\bf z}_A,{\bf z}_A')& = & H_B + {\bf z}_A  \cdot {\bf s}_{N_A+1} 
                      + {\bf z}_A' \cdot {\bf s}_{  N  } ,
\end{eqnarray}
where 
${\bf z}_A  \equiv \langle {\bf s}_{N_A  } \rangle $
and 
${\bf z}_A' \equiv \langle {\bf s}_{    1} \rangle $
are the expectation values obtained from the ground state of $\tilde{H}_A$,
while
${\bf z}_B  \equiv \langle {\bf s}_{N_A+1} \rangle $
and
${\bf z}_B' \equiv \langle {\bf s}_{N    } \rangle $
are derived from the ground state of and $\tilde{H}_B$. \\
\begin{figure}[hptb]
\begin{center}
\includegraphics[width=8.5cm]{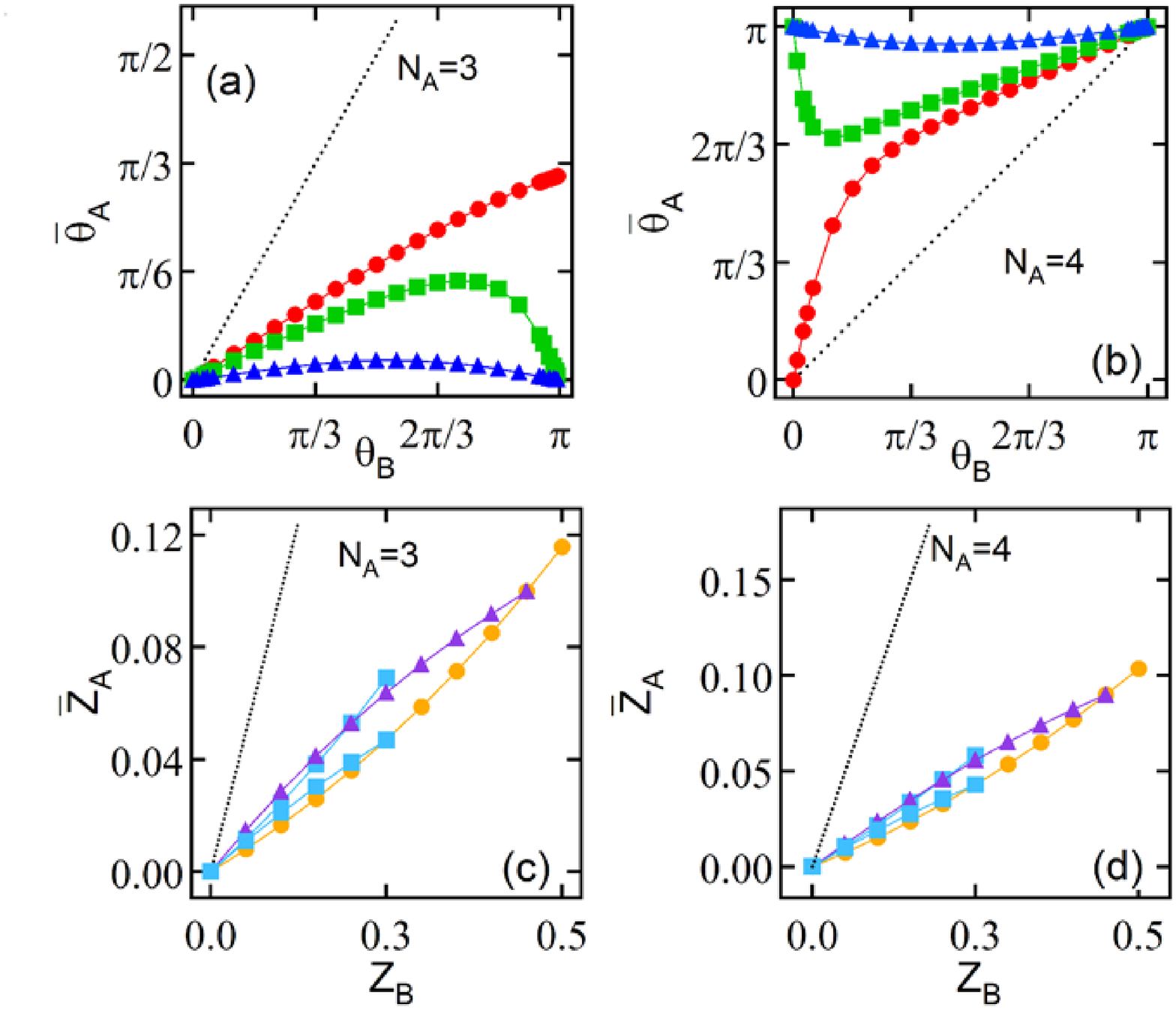}
\end{center}
\caption{(color online) 
(a,b) Relative angle between the boundary spins (${\bar\theta}_A$) as a function of that between the local fields ($\theta_B$), in a $3$- and $4$-qubit chains.
Different symbols refer to different values of the difference in modulus between the fields: $ Z_B = 0 $ (dots), $ 0.25 $ (squares), and $ 0.45 $ (triangles). 
All the results correspond to $ | {\bf z}_B | = 1/2 $, that maximize (minimize) ${\bar\theta}_A$ for $N_A=3$ ($N_A=4$), and for any given value of $Z_B$.  
(c,d) Difference between the moduli of the boundary spins of $A$ (${\bar Z}_A$), as a function of $Z_B$, for relative angles $ {\bar\theta}_A = 0 $ and $ {\bar\theta}_A = \pi $ ($N_A=3$ and $N_A=4$, respectively). 
Different symbols correspond to different values of the moduli:
$ | {\bf z}_B| = 0.5 $ (dots), $ 0.25 $ (squares), and $ 0.05 $ (triangles).}
\label{fig1}
\end{figure}
Intuition suggests that the minimum $E_{bs}(N_A,N_B)$ is achieved for states $ | \Psi_A \rangle $ and $ | \Psi_B \rangle $ that fulfill the following conditions:
$
{\bf z}_A' = (-1)^{N_A+1} {\bf z}_A , \ {\bf z}_B' = (-1)^{N_B+1} {\bf z}_B ,
$
and such that ${\bf z}_A$ and ${\bf z}_B$ (${\bf z}_A'$ and ${\bf z}_B'$) are antiparallel to each other.
More generally, the relevant self-consistent solutions of Eq. \ref{selfcon} can be assumed to fulfil the conditions: 
\begin{equation}\label{intuition}
{\bf z}_A' = \eta {\bf z}_A , \ {\bf z}_B' = \eta {\bf z}_B \ \ (\eta = \pm 1).
\end{equation}
This allows one to reduce the dimensionality of the parameter space where to seek self-consistent solutions. In particular, the set of relevant parameters reduces to 
$ z_A = |{\bf z}_A | = |{\bf z}_A' | $, $ z_B = |{\bf z}_B | = |{\bf z}_B' | $.

\subsubsection{Comparison with not (anti)symmetric subsystem boundary conditions}

In order to support the above conjecture, we consider, in a number of prototypical spin systems, states that don't fulfil the conditions reported in Eq. \ref{intuition}. In this case, the boundary spins of each subsystem differ from one another either in modulus:
\begin{equation}
Z_\alpha \equiv ||{\bf z}_{\alpha}|-|{\bf z}'_{\alpha}|| > 0,
\end{equation}
or in orientation:
\begin{equation}
\cos\theta_{{\alpha}} \equiv \frac{{\bf z}_{\alpha}\cdot{\bf z}'_{\alpha}}{|{\bf z}_{\alpha}||{\bf z}'_{\alpha}|} \neq (-1)^{N_\alpha +1} ,
\end{equation}
or in both respects (with $\alpha=A,B$). 

Within an iterative procedure, given the expectation values of ${\bf s}_{N_A+1}$ and ${\bf s}_{N}$ (${\bf z}_B$ and ${\bf z}_B'$), one can compute the ground state $|\Psi_A\rangle$ of ${\tilde H}_A$, and label the corresponding expectation values of the boundary spins ${\bar{\bf z}}_A$ and ${\bar{\bf z}}_A'$. Analogously, the expectation values derived from the ground state $|\Psi_B\rangle$ of 
$ {\tilde H}_B ( {\bar{\bf z}}_A , {\bar{\bf z}}_A' ) $
are labeled
${\bar{\bar{\bf z}}}_B$ and ${\bar{\bar{\bf z}}}_B'$. 
In summary:
\begin{equation}
(           {\bf z  }_B ,           {\bf z  }_B' )  
\overset{{\tilde H}_A}{\longrightarrow}
(      {\bar{\bf z }}_A ,      {\bar{\bf z }}_A' )  
\overset{{\tilde H}_B}{\longrightarrow}
( {\bar{\bar{\bf z}}}_B , {\bar{\bar{\bf z}}}_B' ) .
\end{equation}
If $|\Psi_A\rangle \otimes |\Psi_B\rangle$ coincides with a self-consistent solution, then
${\bar{\bar{\bf z}}}_B  = {\bf z}_B $ 
and 
${\bar{\bar{\bf z}}}_B' = {\bf z}_B'$.
Correspondingly, $ {\bar{\bar{Z}}}_B = Z_B $ and $ {\bar{\bar{\theta}}}_B = \theta_B $.
Our purpose is to show that the above conditions are not fulfilled by states $ | \Psi_A \rangle \otimes | \Psi_B \rangle $ that violate Eq. \ref{intuition}. 

Let's start by considering the case where the $N$-spin ring is partitioned into two odd-numbered spin segments $A$ and $B$. Here, the above statement follows from the fact that the angle between the expectation values of the boundary spins tends to be smaller than that between the spins with which they interact. As a consequence, for any finite value of $\theta_B$, one has that:
\begin{equation}
\theta_B > {\bar{\theta}}_A > {\bar{\bar{\theta}}}_B
\Rightarrow 
{\bar{\bar{\theta}}}_B \neq \theta_B .
\end{equation}
This is shown in Fig. \ref{fig1}(a), for the case of a three-spin segment, formed by 1/2 spins. The value of $ {\bar{\theta}}_A - \theta_B $ depends on the initial difference between the moduli of the boundary spins ($ Z_B $), and tends to increase with increasing $Z_B$. However, in all the considered cases, the only exception to the case $ {\bar{\theta}}_A < \theta_B $ is found for parallel spins ($ {\bar{\theta}}_A = \theta_B = 0$).
Within such states, we note that the difference in moduli systematically decreases (Fig. \ref{fig1}(c)), which implies, for all finite values of $Z_B$:
\begin{equation}\label{ineq}
Z_B > {\bar{Z}}_A > {\bar{\bar{Z}}}_B
\Rightarrow 
{\bar{\bar{Z}}}_B \neq Z_B .
\end{equation}
Therefore, the possibility of a self-consistent solution implies the condition $ {\bar Z}_A = Z_B = 0 $, as conjectured in Eq. \ref{intuition}. 

In the case where the $N$-spin ring is partitioned into two even-numbered spin segments, the angle between the expectation values of the boundary spins tends to be larger than that between the spins with which they interact. As a consequence,
\begin{equation}
\theta_B < {\bar{\theta}}_A < {\bar{\bar{\theta}}}_B
\Rightarrow 
{\bar{\bar{\theta}}}_B \neq \theta_B .
\end{equation}
This is shown in Fig. \ref{fig1}(c), for the case of a segment formed by four $s=1/2$ spins. Also in this case, the value of $ {\bar{\theta}}_B - \theta_A $ depends monotonically on the initial difference between the moduli of the boundary spins ($ Z_B $). The exceptions to the case $ {\bar{\theta}}_A > \theta_B $ are found both for parallel and antiparallel spins ($ {\bar{\theta}}_A = \theta_B = 0,\pi $). In the latter case (Fig. \ref{fig1}(d)), the inequalities Eq. \ref{ineq} only admit the exception $ {\bar{Z}}_B = Z_B = 0 $. A similar behavior is found in the former case (not shown). Also for partitions in even-numbered spin segments, self-consistent solution thus require the condition Eq. \ref{intuition}. 

\begin{figure}[hptb]
\begin{center}
\includegraphics[width=8.5cm]{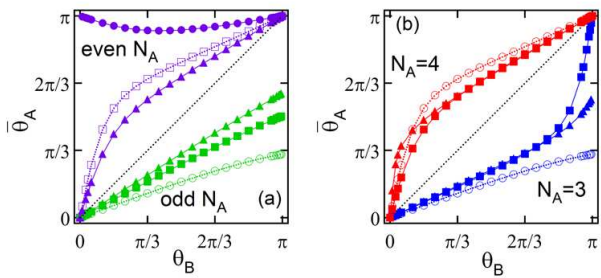}
\end{center}
\caption{(color online) 
(a) Relative angle of the boundary spins of $A$ (${\bar\theta}_A$) as a function of $\theta_B$ for different lengths $N_A$ of the qubit chain. Green corresponds to odd-numbered chains ($N_A=2n+1$), while violet refers to even-numbered ones ($N_A=2n$). Different symbols correspond to different values of $n$: $1$ (dots), $2$ (squares), $3$ (triangles). (b) Angle ${\bar\theta}_A$ as a function of $\theta_B$, for chains formed by spins $ s > 1/2 $. Red and blue correspond to 4- and 3-spin chains, respectively. Different symbols correspond to different spin lengths: $s=1/2$ (dots), $1$ (squares), $3/2$ (triangles). All the reported values are obtained for $ |{\bf z}_B| = s $ and $ Z_B = 0 $, which maximize (minimize) ${\bar\theta}_A$, for each given value of $\theta_B$.}
\label{fig1bis}
\end{figure}

These features persist for different lengths of the chains (Fig. \ref{fig1bis} (a)). Irrespective of the initial difference in moduli ($Z_B$), parallel-oriented boundary spins ($\theta_B=0$) are the ones that can allow a self-consistent solution ($ {\bar{\theta}}_A = \theta_B$) in the case of odd-numbered chains.
Within the considered even-numbered chains, both the parallel and - with the exception of the 2-qubit subsystem - antiparallel orientations can allow a self-consistent solution. 
We note that, for any initial condition 
$ (           {\bf z  }_B ,           {\bf z  }_B' ) $,
the difference between $\theta_B$ and ${\bar\theta}_A$ tends to decrease with increasing number of spins. In all the considered cases, only equal moduli ($Z_B=0$) and antiparallel orientations ($\theta_B=\pi$) provide a self-consistent solution of Eq. \ref{selfcon}. 

We finally consider the dependence on the length of the spins (Fig. \ref{fig1bis} (b)). The main features of even-numbered qubit chains are preserved when $s>1/2$. Odd-numbered chains present instead a different behavior for integer and half-integer spins: in the former case, only parallel orientations of the boundary spins can allow a self-consistent solution, whereas in the latter case we find also $ \theta_B = \pi $. In all cases, the condition $Z_B=0$ is required.

\subsubsection{Energy minima for biseparable states}

\begin{figure}[ptb]
\begin{center}
\includegraphics[width=8.5cm]{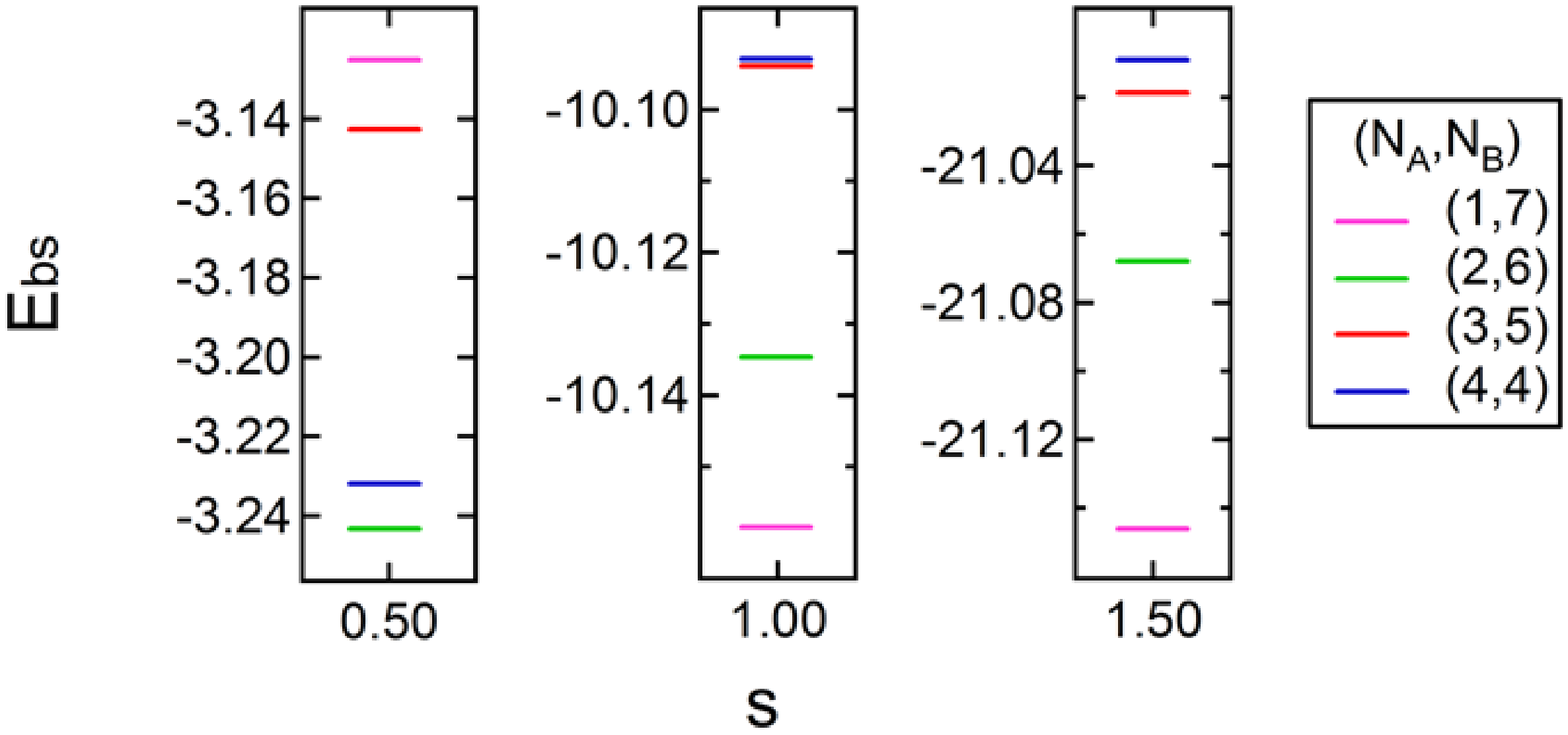}
\end{center}
\caption{(color online) 
Values of the energy minima $E_{bs}(N_A,N_B)$ calculated for the different bipartitions of the $N=8$ spin ring, and for spin length $ s = 1/2 , 1 , 3/2 $. 
}
\label{fig2}
\end{figure}

Having shown that the search of a self consistent solution can be restricted to boundary spins having the same modulus, and parallel or antiparallel orientation, we now seek the energy minima corresponding to biseparable states. In particular, we consider different bipartitions of a ring formed by $N=8$ spins $s$, with $s=1/2,1,3/2$. The corresponding energy minima are displayed in Fig. \ref{fig2}. 
As in the spin chain \cite{troiani2012energy}, the overall energy minimum is achieved within the partition $ \left(N_A,N_B \right) = (2,6) $ for the qubit case. For $s>1/2$, instead, it corresponds to $ \left(N_A,N_B \right) = (1,7) $. 
We also note that, in the qubit ring, the minima $ E_{bs} (N_A,N_B) $ with even values of $N_A$ and $N_B$ correspond to states with $ \langle H_{A,B} \rangle = 0 $, i.e. with vanishing expectation values of the edge spins ${\bf s}_1$, ${\bf s}_{N_A}$, ${\bf s}_{N_A+1}$, and ${\bf s}_N$. In all the other cases ($ s > 1/2 $), $ \langle H_{A,B} \rangle < 0 $ for the lowest-energy biseparable states. 

\section{Energy as a witness of entanglement between each spin and rest of the ring}
\label{sec2}
The rings considered so far are all formed by equivalent spins. Such equivalence breaks down if a magnetic defect ($s_M \neq s$) is introduced in the ring, as is the case with heterometallic wheels. The spin chain can also be regarded as a particular example of this class of systems, corresponding to the case where the defect is spinless ($s_M=0$). 
Here, the expectation value of the exchange Hamiltonian averages over inequivalent contributions, and cannot provide information on the entanglement properties of each specific pair. The local access to single exchange operators is thus required in order to selectively detect entanglement between each given spin pair \cite{siloi2012towards}. 

Hereafter we show that, in spite of its nonlocal character, the witness $H$ allows the selective detection of entanglement between each (inequivalent) spin and all the others. To this aim, we derive energy minima 
\begin{equation}
E_{bs}^k \equiv E_{bs}^k (N_A=1,N_B=N-1)
\end{equation} 
corresponding to the absence of entanglement between any given spin ${\bf s}_k$ and the rest of the system. The minimum $E_{bs}^k$ thus refers to a particular bipartition of the system, where the subsystem $A$ is formed uniquely by the $k-$th spin. Its value coincides with the ground state energy of 
\begin{equation}
\tilde{H}_B (z_A=s_k) = H_B + s_k (s_{z,k-1}+s_{z,k+1}) 
\end{equation}
(where $H_B$ is given by Eq. \ref{HABa}), which corresponds to the Hamiltonian of an open spin chain with a magnetic field $s_k$ locally applied at the edge spins. The above value of $z_A$ results from the ground state of 
$ \tilde{H}_A = s_{z,k} (z_B +z_B') $, 
which trivially corresponds to $m_{k}=s_k$, if the orientation of $\hat{\bf z}$ is defined such that $ z_B + z_B' < 0 $.

In the following, we apply the above approach to a class of spin models, whose physical implementation is represented by the series of Cr$_7$M molecular nanomagnets \cite{Caciuffo05}. In these molecular spin clusters, the magnetic defect is represented by the metal $M$, which replaces one of the eight Cr ions in the parent Cr$_8$ molecule. In particular, the chemical substitutions are: 
M = Zn, Cu, Ni, Cr, Fe, and Mn. These ions carry spins: $s_M = 0, 1/2, 1, 3/2, 2,$ and $ 5/2$, respectively. In all these nanomagnets, the dominant part of the spin Hamiltonian, to which we shall limit ourselves hereafter, is represented by isotropic exchange. Besides, the values of the exchange couplings are substantially identical for all the spin pairs, independently from the particular chemical substitution. 

\begin{figure}[tb]
\begin{center}
\includegraphics[width=8.5cm]{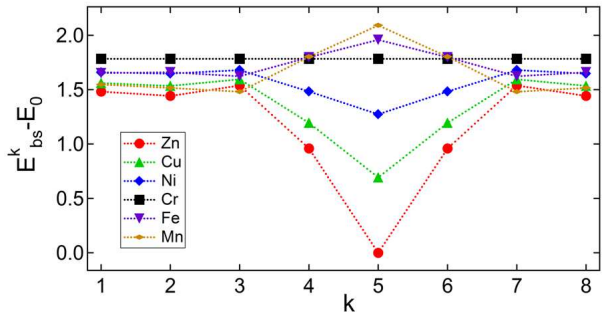}
\end{center}
\caption{(color online) 
Energetic cost $E_{bs}^k-E_{0}$ of disentangling the $k-$th spin from the remaining ones, for the different molecules of the Cr$_7$M series. The position of the substituted ion corresponds to $k=5$. }
\label{fig3}
\end{figure}

The values of the minima $E_{bs}^k$ for the different molecules, referred to the respective ground state energies $E_{0}$, are reported in Fig. \ref{fig3}. 
The quantity $E_{bs}^k-E_{0}$ can be interpreted as the minimum energy required to disentangle the $k-$th ion from the rest of the system, starting from its ground state. 
While such energetic cost is independent on the position of the spin in the Cr$_8$ case (black squares), it becomes spatially modulated as a result of the chemical substitution. 
The dependence of $E_{bs}^k-E_{0}$ on the spin defect $s_M$ is maximum for the substituted spin ($k=5$) and for the neighboring ones ($k=4,6$). 
In particular, the value of $E_{bs}^5-E_{0}$ increases monotonically with the length of the spin defect, and so does the gap corresponding to the disentangling of its neighboring spins. 

For the systems with $s_M < s_{Cr} = 3/2$, an expectation value of the exchange Hamiltonian that fulfills the inequality 
$
\langle H \rangle < \min_k \{ E^k_{bs} \} = E^5_{bs}
$
implies that each of the spins is entangled with the rest of the system. Expectation values such that 
$
E^5_{bs} \le \langle H \rangle < E^4_{bs}=E^6_{bs}
$
allow one to draw the same conclusion for all the spins but $s_5$. Analogously, larger and larger values of energy provide information on fewer and fewer spins, until no conclusion can be drawn for 
$
\langle H \rangle \ge \max_k \{ E^k_{bs} \} 
$. 
For the systems with $s_M > s_{Cr} = 3/2$, the ordering of the thresholds $E^k_{bs}$ is approximately inverted, such that the presence of entanglement between $s_M$ and the rest of the spins can be detected in the largest energy and temperature ranges.

\section{Conclusions}

In conclusion, we have proved that the ground state of an $N$-spin ring, with antiferromagnetic exchange between nearest neighbors, is characterized by $N$-partite entanglement, and that such property is also present in the low-temperature equilibrium state. We have developed a general approach for deriving energy minima corresponding to biseparable states. These allows one to derive inequalities, whose violation enables the detection of multipartite entanglement. Along the same lines, we have shown that exchange energy can be used to infer the presence of entanglement between individual spins and the rest of the system. The related energy minima have been derived for a class of heterometallic, ring-shaped molecular nanomagnets. 


\begin{thebibliography}{51}
\expandafter\ifx\csname natexlab\endcsname\relax\def\natexlab#1{#1}\fi
\expandafter\ifx\csname bibnamefont\endcsname\relax
  \def\bibnamefont#1{#1}\fi
\expandafter\ifx\csname bibfnamefont\endcsname\relax
  \def\bibfnamefont#1{#1}\fi
\expandafter\ifx\csname citenamefont\endcsname\relax
  \def\citenamefont#1{#1}\fi
\expandafter\ifx\csname url\endcsname\relax
  \def\url#1{\texttt{#1}}\fi
\expandafter\ifx\csname urlprefix\endcsname\relax\def\urlprefix{URL }\fi
\providecommand{\bibinfo}[2]{#2}
\providecommand{\eprint}[2][]{\url{#2}}

\bibitem[{\citenamefont{Nielsen and Chuang}(2000)}]{nielsen_chuang01}
\bibinfo{author}{\bibfnamefont{M.~A.} \bibnamefont{Nielsen}} \bibnamefont{and}
  \bibinfo{author}{\bibfnamefont{I.~L.} \bibnamefont{Chuang}},
  \emph{\bibinfo{title}{Quantum Computation and quantum information}}
  (\bibinfo{publisher}{Cambridge University Press},
  \bibinfo{address}{Cambridge}, \bibinfo{year}{2000}).

\bibitem[{\citenamefont{Horodecki et~al.}(2009)\citenamefont{Horodecki,
  Horodecki, Horodecki, and Horodecki}}]{Hor+09rmp}
\bibinfo{author}{\bibfnamefont{R.}~\bibnamefont{Horodecki}},
  \bibinfo{author}{\bibfnamefont{P.}~\bibnamefont{Horodecki}},
  \bibinfo{author}{\bibfnamefont{M.}~\bibnamefont{Horodecki}},
  \bibnamefont{and}
  \bibinfo{author}{\bibfnamefont{K.}~\bibnamefont{Horodecki}},
  \bibinfo{journal}{Rev. Mod. Phys.} \textbf{\bibinfo{volume}{81}},
  \bibinfo{pages}{865} (\bibinfo{year}{2009}).

\bibitem[{\citenamefont{G{\"u}hne and T{\'o}th}(2009)}]{guhne2009entanglement}
\bibinfo{author}{\bibfnamefont{O.}~\bibnamefont{G{\"u}hne}} \bibnamefont{and}
  \bibinfo{author}{\bibfnamefont{G.}~\bibnamefont{T{\'o}th}},
  \bibinfo{journal}{Phys. Rep.} \textbf{\bibinfo{volume}{474}},
  \bibinfo{pages}{1} (\bibinfo{year}{2009}).

\bibitem[{\citenamefont{Kiesel et~al.}(2007)\citenamefont{Kiesel, Schmid,
  T\'oth, Solano, and Weinfurter}}]{kiesel2007experimental}
\bibinfo{author}{\bibfnamefont{N.}~\bibnamefont{Kiesel}},
  \bibinfo{author}{\bibfnamefont{C.}~\bibnamefont{Schmid}},
  \bibinfo{author}{\bibfnamefont{G.}~\bibnamefont{T\'oth}},
  \bibinfo{author}{\bibfnamefont{E.}~\bibnamefont{Solano}}, \bibnamefont{and}
  \bibinfo{author}{\bibfnamefont{H.}~\bibnamefont{Weinfurter}},
  \bibinfo{journal}{Phys. Rev. Lett.} \textbf{\bibinfo{volume}{98}},
  \bibinfo{pages}{063604} (\bibinfo{year}{2007}).

\bibitem[{\citenamefont{Wieczorek et~al.}(2009)\citenamefont{Wieczorek,
  Krischek, Kiesel, Michelberger, T{\'o}th, and
  Weinfurter}}]{wieczorek2009experimental}
\bibinfo{author}{\bibfnamefont{W.}~\bibnamefont{Wieczorek}},
  \bibinfo{author}{\bibfnamefont{R.}~\bibnamefont{Krischek}},
  \bibinfo{author}{\bibfnamefont{N.}~\bibnamefont{Kiesel}},
  \bibinfo{author}{\bibfnamefont{P.}~\bibnamefont{Michelberger}},
  \bibinfo{author}{\bibfnamefont{G.}~\bibnamefont{T{\'o}th}}, \bibnamefont{and}
  \bibinfo{author}{\bibfnamefont{H.}~\bibnamefont{Weinfurter}},
  \bibinfo{journal}{Phys. Rev. Lett.} \textbf{\bibinfo{volume}{103}},
  \bibinfo{pages}{020504} (\bibinfo{year}{2009}).

\bibitem[{\citenamefont{Lu et~al.}(2007)\citenamefont{Lu, Zhou, G{\"u}hne, Gao,
  Zhang, Yuan, Goebel, Yang, and Pan}}]{lu2007experimental}
\bibinfo{author}{\bibfnamefont{C.-Y.} \bibnamefont{Lu}},
  \bibinfo{author}{\bibfnamefont{X.-Q.} \bibnamefont{Zhou}},
  \bibinfo{author}{\bibfnamefont{O.}~\bibnamefont{G{\"u}hne}},
  \bibinfo{author}{\bibfnamefont{W.-B.} \bibnamefont{Gao}},
  \bibinfo{author}{\bibfnamefont{J.}~\bibnamefont{Zhang}},
  \bibinfo{author}{\bibfnamefont{Z.-S.} \bibnamefont{Yuan}},
  \bibinfo{author}{\bibfnamefont{A.}~\bibnamefont{Goebel}},
  \bibinfo{author}{\bibfnamefont{T.}~\bibnamefont{Yang}}, \bibnamefont{and}
  \bibinfo{author}{\bibfnamefont{J.-W.} \bibnamefont{Pan}},
  \bibinfo{journal}{Nat. Phys.} \textbf{\bibinfo{volume}{3}},
  \bibinfo{pages}{91} (\bibinfo{year}{2007}).

\bibitem[{\citenamefont{Monz et~al.}(2011)\citenamefont{Monz, Schindler,
  Barreiro, Chwalla, Nigg, Coish, Harlander, H\"ansel, Hennrich, and
  Blatt}}]{monz2011qubit}
\bibinfo{author}{\bibfnamefont{T.}~\bibnamefont{Monz}},
  \bibinfo{author}{\bibfnamefont{P.}~\bibnamefont{Schindler}},
  \bibinfo{author}{\bibfnamefont{J.~T.} \bibnamefont{Barreiro}},
  \bibinfo{author}{\bibfnamefont{M.}~\bibnamefont{Chwalla}},
  \bibinfo{author}{\bibfnamefont{D.}~\bibnamefont{Nigg}},
  \bibinfo{author}{\bibfnamefont{W.~A.} \bibnamefont{Coish}},
  \bibinfo{author}{\bibfnamefont{M.}~\bibnamefont{Harlander}},
  \bibinfo{author}{\bibfnamefont{W.}~\bibnamefont{H\"ansel}},
  \bibinfo{author}{\bibfnamefont{M.}~\bibnamefont{Hennrich}}, \bibnamefont{and}
  \bibinfo{author}{\bibfnamefont{R.}~\bibnamefont{Blatt}},
  \bibinfo{journal}{Phys. Rev. Lett.} \textbf{\bibinfo{volume}{106}},
  \bibinfo{pages}{130506} (\bibinfo{year}{2011}).

\bibitem[{\citenamefont{Hammerer et~al.}(2010)\citenamefont{Hammerer,
  S\o{}rensen, and Polzik}}]{hammerer2010rev}
\bibinfo{author}{\bibfnamefont{K.}~\bibnamefont{Hammerer}},
  \bibinfo{author}{\bibfnamefont{A.~S.} \bibnamefont{S\o{}rensen}},
  \bibnamefont{and} \bibinfo{author}{\bibfnamefont{E.~S.}
  \bibnamefont{Polzik}}, \bibinfo{journal}{Rev. Mod. Phys.}
  \textbf{\bibinfo{volume}{82}}, \bibinfo{pages}{1041} (\bibinfo{year}{2010}).

\bibitem[{\citenamefont{Neumann et~al.}(2008)\citenamefont{Neumann, Mizuochi,
  Rempp, Hemmer, Watanabe, Yamasaki, Jacques, Gaebel, Jelezko, and
  Wrachtrup}}]{neumann2008multipartite}
\bibinfo{author}{\bibfnamefont{P.}~\bibnamefont{Neumann}},
  \bibinfo{author}{\bibfnamefont{N.}~\bibnamefont{Mizuochi}},
  \bibinfo{author}{\bibfnamefont{F.}~\bibnamefont{Rempp}},
  \bibinfo{author}{\bibfnamefont{P.}~\bibnamefont{Hemmer}},
  \bibinfo{author}{\bibfnamefont{H.}~\bibnamefont{Watanabe}},
  \bibinfo{author}{\bibfnamefont{S.}~\bibnamefont{Yamasaki}},
  \bibinfo{author}{\bibfnamefont{V.}~\bibnamefont{Jacques}},
  \bibinfo{author}{\bibfnamefont{T.}~\bibnamefont{Gaebel}},
  \bibinfo{author}{\bibfnamefont{F.}~\bibnamefont{Jelezko}}, \bibnamefont{and}
  \bibinfo{author}{\bibfnamefont{J.}~\bibnamefont{Wrachtrup}},
  \bibinfo{journal}{Science} \textbf{\bibinfo{volume}{320}},
  \bibinfo{pages}{1326} (\bibinfo{year}{2008}).

\bibitem[{\citenamefont{Raussendorf et~al.}(2003)\citenamefont{Raussendorf,
  Browne, and Briegel}}]{Raussendorf}
\bibinfo{author}{\bibfnamefont{R.}~\bibnamefont{Raussendorf}},
  \bibinfo{author}{\bibfnamefont{D.~E.} \bibnamefont{Browne}},
  \bibnamefont{and} \bibinfo{author}{\bibfnamefont{H.~J.}
  \bibnamefont{Briegel}}, \bibinfo{journal}{Phys. Rev. A}
  \textbf{\bibinfo{volume}{68}}, \bibinfo{pages}{022312}
  (\bibinfo{year}{2003}).

\bibitem[{\citenamefont{Briegel et~al.}(2009)\citenamefont{Briegel, Browne,
  D{\"u}r, Raussendorf, and Van~den Nest}}]{briegel2009measurement}
\bibinfo{author}{\bibfnamefont{H.}~\bibnamefont{Briegel}},
  \bibinfo{author}{\bibfnamefont{D.}~\bibnamefont{Browne}},
  \bibinfo{author}{\bibfnamefont{W.}~\bibnamefont{D{\"u}r}},
  \bibinfo{author}{\bibfnamefont{R.}~\bibnamefont{Raussendorf}},
  \bibnamefont{and} \bibinfo{author}{\bibfnamefont{M.}~\bibnamefont{Van~den
  Nest}}, \bibinfo{journal}{Nat. Phys.} \textbf{\bibinfo{volume}{5}},
  \bibinfo{pages}{19} (\bibinfo{year}{2009}).

\bibitem[{\citenamefont{Giovannetti et~al.}(2004)\citenamefont{Giovannetti,
  Lloyd, and Maccone}}]{giovannetti2004quantum}
\bibinfo{author}{\bibfnamefont{V.}~\bibnamefont{Giovannetti}},
  \bibinfo{author}{\bibfnamefont{S.}~\bibnamefont{Lloyd}}, \bibnamefont{and}
  \bibinfo{author}{\bibfnamefont{L.}~\bibnamefont{Maccone}},
  \bibinfo{journal}{Science} \textbf{\bibinfo{volume}{306}},
  \bibinfo{pages}{1330} (\bibinfo{year}{2004}).

\bibitem[{\citenamefont{S{\o}rensen et~al.}(2001)\citenamefont{S{\o}rensen,
  Duan, Cirac, and Zoller}}]{sorensen2001many}
\bibinfo{author}{\bibfnamefont{A.}~\bibnamefont{S{\o}rensen}},
  \bibinfo{author}{\bibfnamefont{L.-M.} \bibnamefont{Duan}},
  \bibinfo{author}{\bibfnamefont{J.}~\bibnamefont{Cirac}}, \bibnamefont{and}
  \bibinfo{author}{\bibfnamefont{P.}~\bibnamefont{Zoller}},
  \bibinfo{journal}{Nature} \textbf{\bibinfo{volume}{409}}, \bibinfo{pages}{63}
  (\bibinfo{year}{2001}).

\bibitem[{\citenamefont{Hyllus et~al.}(2012)\citenamefont{Hyllus, Laskowski,
  Krischek, Schwemmer, Wieczorek, Weinfurter, Pezz\'e, and
  Smerzi}}]{hyllus2012fisher}
\bibinfo{author}{\bibfnamefont{P.}~\bibnamefont{Hyllus}},
  \bibinfo{author}{\bibfnamefont{W.}~\bibnamefont{Laskowski}},
  \bibinfo{author}{\bibfnamefont{R.}~\bibnamefont{Krischek}},
  \bibinfo{author}{\bibfnamefont{C.}~\bibnamefont{Schwemmer}},
  \bibinfo{author}{\bibfnamefont{W.}~\bibnamefont{Wieczorek}},
  \bibinfo{author}{\bibfnamefont{H.}~\bibnamefont{Weinfurter}},
  \bibinfo{author}{\bibfnamefont{L.}~\bibnamefont{Pezz\'e}}, \bibnamefont{and}
  \bibinfo{author}{\bibfnamefont{A.}~\bibnamefont{Smerzi}},
  \bibinfo{journal}{Phys. Rev. A} \textbf{\bibinfo{volume}{85}},
  \bibinfo{pages}{022321} (\bibinfo{year}{2012}).

\bibitem[{\citenamefont{T\'oth}(2012)}]{toth2012multipartite}
\bibinfo{author}{\bibfnamefont{G.}~\bibnamefont{T\'oth}},
  \bibinfo{journal}{Phys. Rev. A} \textbf{\bibinfo{volume}{85}},
  \bibinfo{pages}{022322} (\bibinfo{year}{2012}).

\bibitem[{\citenamefont{Amico et~al.}(2008)\citenamefont{Amico, Fazio,
  Osterloh, and Vedral}}]{Amc+08rmp}
\bibinfo{author}{\bibfnamefont{L.}~\bibnamefont{Amico}},
  \bibinfo{author}{\bibfnamefont{R.}~\bibnamefont{Fazio}},
  \bibinfo{author}{\bibfnamefont{A.}~\bibnamefont{Osterloh}}, \bibnamefont{and}
  \bibinfo{author}{\bibfnamefont{V.}~\bibnamefont{Vedral}},
  \bibinfo{journal}{Rev. Mod. Phys.} \textbf{\bibinfo{volume}{80}},
  \bibinfo{pages}{517} (\bibinfo{year}{2008}).

\bibitem[{\citenamefont{Wang}(2002)}]{wan02pra}
\bibinfo{author}{\bibfnamefont{X.}~\bibnamefont{Wang}}, \bibinfo{journal}{Phys.
  Rev. A} \textbf{\bibinfo{volume}{66}}, \bibinfo{pages}{044305}
  (\bibinfo{year}{2002}).

\bibitem[{\citenamefont{Bru{\ss} et~al.}(2005)\citenamefont{Bru{\ss}, Datta,
  Ekert, Kwek, and Macchiavello}}]{bruss2005multipartite}
\bibinfo{author}{\bibfnamefont{D.}~\bibnamefont{Bru{\ss}}},
  \bibinfo{author}{\bibfnamefont{N.}~\bibnamefont{Datta}},
  \bibinfo{author}{\bibfnamefont{A.}~\bibnamefont{Ekert}},
  \bibinfo{author}{\bibfnamefont{L.~C.} \bibnamefont{Kwek}}, \bibnamefont{and}
  \bibinfo{author}{\bibfnamefont{C.}~\bibnamefont{Macchiavello}},
  \bibinfo{journal}{Phys. Rev. A} \textbf{\bibinfo{volume}{72}},
  \bibinfo{pages}{014301} (\bibinfo{year}{2005}).

\bibitem[{\citenamefont{G{\"u}hne and T{\'o}th}(2006)}]{guhne_toth06pra}
\bibinfo{author}{\bibfnamefont{O.}~\bibnamefont{G{\"u}hne}} \bibnamefont{and}
  \bibinfo{author}{\bibfnamefont{G.}~\bibnamefont{T{\'o}th}},
  \bibinfo{journal}{Physical Review A} \textbf{\bibinfo{volume}{73}},
  \bibinfo{pages}{052319} (\bibinfo{year}{2006}).

\bibitem[{\citenamefont{Troiani}(2011)}]{troiani2011entanglement}
\bibinfo{author}{\bibfnamefont{F.}~\bibnamefont{Troiani}},
  \bibinfo{journal}{Phys. Rev. A} \textbf{\bibinfo{volume}{83}},
  \bibinfo{pages}{022324} (\bibinfo{year}{2011}).

\bibitem[{\citenamefont{Terhal}(2000)}]{terhal2000bell}
\bibinfo{author}{\bibfnamefont{B.~M.} \bibnamefont{Terhal}},
  \bibinfo{journal}{Phys. Lett. A} \textbf{\bibinfo{volume}{271}},
  \bibinfo{pages}{319} (\bibinfo{year}{2000}).

\bibitem[{\citenamefont{Lewenstein et~al.}(2000)\citenamefont{Lewenstein,
  Kraus, Cirac, and Horodecki}}]{lewenstein2000optimization}
\bibinfo{author}{\bibfnamefont{M.}~\bibnamefont{Lewenstein}},
  \bibinfo{author}{\bibfnamefont{B.}~\bibnamefont{Kraus}},
  \bibinfo{author}{\bibfnamefont{J.~I.} \bibnamefont{Cirac}}, \bibnamefont{and}
  \bibinfo{author}{\bibfnamefont{P.}~\bibnamefont{Horodecki}},
  \bibinfo{journal}{Phys. Rev. A} \textbf{\bibinfo{volume}{62}},
  \bibinfo{pages}{052310} (\bibinfo{year}{2000}).

\bibitem[{\citenamefont{Ac\'in et~al.}(2001)\citenamefont{Ac\'in, Bru\ss{},
  Lewenstein, and Sanpera}}]{acin2001classification}
\bibinfo{author}{\bibfnamefont{A.}~\bibnamefont{Ac\'in}},
  \bibinfo{author}{\bibfnamefont{D.}~\bibnamefont{Bru\ss{}}},
  \bibinfo{author}{\bibfnamefont{M.}~\bibnamefont{Lewenstein}},
  \bibnamefont{and} \bibinfo{author}{\bibfnamefont{A.}~\bibnamefont{Sanpera}},
  \bibinfo{journal}{Phys. Rev. Lett.} \textbf{\bibinfo{volume}{87}},
  \bibinfo{pages}{040401} (\bibinfo{year}{2001}).

\bibitem[{\citenamefont{Bourennane et~al.}(2004)\citenamefont{Bourennane, Eibl,
  Kurtsiefer, Gaertner, Weinfurter, G\"uhne, Hyllus, Bru\ss{}, Lewenstein, and
  Sanpera}}]{bourennane2004experimental}
\bibinfo{author}{\bibfnamefont{M.}~\bibnamefont{Bourennane}},
  \bibinfo{author}{\bibfnamefont{M.}~\bibnamefont{Eibl}},
  \bibinfo{author}{\bibfnamefont{C.}~\bibnamefont{Kurtsiefer}},
  \bibinfo{author}{\bibfnamefont{S.}~\bibnamefont{Gaertner}},
  \bibinfo{author}{\bibfnamefont{H.}~\bibnamefont{Weinfurter}},
  \bibinfo{author}{\bibfnamefont{O.}~\bibnamefont{G\"uhne}},
  \bibinfo{author}{\bibfnamefont{P.}~\bibnamefont{Hyllus}},
  \bibinfo{author}{\bibfnamefont{D.}~\bibnamefont{Bru\ss{}}},
  \bibinfo{author}{\bibfnamefont{M.}~\bibnamefont{Lewenstein}},
  \bibnamefont{and} \bibinfo{author}{\bibfnamefont{A.}~\bibnamefont{Sanpera}},
  \bibinfo{journal}{Phys. Rev. Lett.} \textbf{\bibinfo{volume}{92}},
  \bibinfo{pages}{087902} (\bibinfo{year}{2004}).

\bibitem[{\citenamefont{Huber et~al.}(2010)\citenamefont{Huber, Mintert,
  Gabriel, and Hiesmayr}}]{huber2010detection}
\bibinfo{author}{\bibfnamefont{M.}~\bibnamefont{Huber}},
  \bibinfo{author}{\bibfnamefont{F.}~\bibnamefont{Mintert}},
  \bibinfo{author}{\bibfnamefont{A.}~\bibnamefont{Gabriel}}, \bibnamefont{and}
  \bibinfo{author}{\bibfnamefont{B.~C.} \bibnamefont{Hiesmayr}},
  \bibinfo{journal}{Phys. Rev. Lett.} \textbf{\bibinfo{volume}{104}},
  \bibinfo{pages}{210501} (\bibinfo{year}{2010}).

\bibitem[{\citenamefont{Duan}(2011)}]{duan2011entanglement}
\bibinfo{author}{\bibfnamefont{L.-M.} \bibnamefont{Duan}},
  \bibinfo{journal}{Phys. Rev. Lett.} \textbf{\bibinfo{volume}{107}},
  \bibinfo{pages}{180502} (\bibinfo{year}{2011}).

\bibitem[{\citenamefont{Sperling and Vogel}(2013)}]{sperling2013multipartite}
\bibinfo{author}{\bibfnamefont{J.}~\bibnamefont{Sperling}} \bibnamefont{and}
  \bibinfo{author}{\bibfnamefont{W.}~\bibnamefont{Vogel}},
  \bibinfo{journal}{Phys. Rev. Lett.} \textbf{\bibinfo{volume}{111}},
  \bibinfo{pages}{110503} (\bibinfo{year}{2013}).

\bibitem[{\citenamefont{Krammer et~al.}(2009)\citenamefont{Krammer, Kampermann,
  Bru\ss{}, Bertlmann, Kwek, and Macchiavello}}]{krammer2009multipartite}
\bibinfo{author}{\bibfnamefont{P.}~\bibnamefont{Krammer}},
  \bibinfo{author}{\bibfnamefont{H.}~\bibnamefont{Kampermann}},
  \bibinfo{author}{\bibfnamefont{D.}~\bibnamefont{Bru\ss{}}},
  \bibinfo{author}{\bibfnamefont{R.~A.} \bibnamefont{Bertlmann}},
  \bibinfo{author}{\bibfnamefont{L.~C.} \bibnamefont{Kwek}}, \bibnamefont{and}
  \bibinfo{author}{\bibfnamefont{C.}~\bibnamefont{Macchiavello}},
  \bibinfo{journal}{Phys. Rev. Lett.} \textbf{\bibinfo{volume}{103}},
  \bibinfo{pages}{100502} (\bibinfo{year}{2009}).

\bibitem[{\citenamefont{Cramer et~al.}(2011)\citenamefont{Cramer, Plenio, and
  Wunderlich}}]{cramer2011measuring}
\bibinfo{author}{\bibfnamefont{M.}~\bibnamefont{Cramer}},
  \bibinfo{author}{\bibfnamefont{M.~B.}~\bibnamefont{Plenio}}, \bibnamefont{and}
  \bibinfo{author}{\bibfnamefont{H.}~\bibnamefont{Wunderlich}},
  \bibinfo{journal}{Phys. Rev. Lett.} \textbf{\bibinfo{volume}{106}},
  \bibinfo{pages}{020401} (\bibinfo{year}{2011}).

\bibitem[{\citenamefont{Brukner et~al.}(2006)\citenamefont{Brukner, Vedral, and
  Zeilinger}}]{brukner2006crucial}
\bibinfo{author}{\bibfnamefont{{\v{C}}.}~\bibnamefont{Brukner}},
  \bibinfo{author}{\bibfnamefont{V.}~\bibnamefont{Vedral}}, \bibnamefont{and}
  \bibinfo{author}{\bibfnamefont{A.}~\bibnamefont{Zeilinger}},
  \bibinfo{journal}{Phys. Rev. A} \textbf{\bibinfo{volume}{73}},
  \bibinfo{pages}{012110} (\bibinfo{year}{2006}).

\bibitem[{\citenamefont{Wie{\'s}niak et~al.}(2005)\citenamefont{Wie{\'s}niak,
  Vedral, and Brukner}}]{wiesniak2005magnetic}
\bibinfo{author}{\bibfnamefont{M.}~\bibnamefont{Wie{\'s}niak}},
  \bibinfo{author}{\bibfnamefont{V.}~\bibnamefont{Vedral}}, \bibnamefont{and}
  \bibinfo{author}{\bibfnamefont{{\v{C}}.}~\bibnamefont{Brukner}},
  \bibinfo{journal}{New Journal of Physics} \textbf{\bibinfo{volume}{7}},
  \bibinfo{pages}{258} (\bibinfo{year}{2005}).

\bibitem[{\citenamefont{Brukner and Vedral}(2004)}]{brukner2004macroscopic}
\bibinfo{author}{\bibfnamefont{{\v{C}}.}~\bibnamefont{Brukner}}
  \bibnamefont{and} \bibinfo{author}{\bibfnamefont{V.}~\bibnamefont{Vedral}},
  \bibinfo{journal}{arXiv preprint quant-ph/0406040}  (\bibinfo{year}{2004}).

\bibitem[{\citenamefont{Wang and Zanardi}(2002)}]{wang2002quantum}
\bibinfo{author}{\bibfnamefont{X.}~\bibnamefont{Wang}} \bibnamefont{and}
  \bibinfo{author}{\bibfnamefont{P.}~\bibnamefont{Zanardi}},
  \bibinfo{journal}{Phys. Lett. A} \textbf{\bibinfo{volume}{301}},
  \bibinfo{pages}{1} (\bibinfo{year}{2002}).

\bibitem[{\citenamefont{S\o{}rensen and
  M\o{}lmer}(2001)}]{sorensen2001entanglement}
\bibinfo{author}{\bibfnamefont{A.~S.} \bibnamefont{S\o{}rensen}}
  \bibnamefont{and}
  \bibinfo{author}{\bibfnamefont{K.}~\bibnamefont{M\o{}lmer}},
  \bibinfo{journal}{Phys. Rev. Lett.} \textbf{\bibinfo{volume}{86}},
  \bibinfo{pages}{4431} (\bibinfo{year}{2001}).

\bibitem[{\citenamefont{Korbicz et~al.}(2005)\citenamefont{Korbicz, Cirac, and
  Lewenstein}}]{korbicz2005spin}
\bibinfo{author}{\bibfnamefont{J.~K.}~\bibnamefont{Korbicz}},
  \bibinfo{author}{\bibfnamefont{J.~I.}~\bibnamefont{Cirac}}, \bibnamefont{and}
  \bibinfo{author}{\bibfnamefont{M.}~\bibnamefont{Lewenstein}},
  \bibinfo{journal}{Phys. Rev. Lett.} \textbf{\bibinfo{volume}{95}},
  \bibinfo{pages}{120502} (\bibinfo{year}{2005}).

\bibitem[{\citenamefont{T\'oth et~al.}(2007)\citenamefont{T\'oth, Knapp,
  G\"uhne, and Briegel}}]{toth2007optimal}
\bibinfo{author}{\bibfnamefont{G.}~\bibnamefont{T\'oth}},
  \bibinfo{author}{\bibfnamefont{C.}~\bibnamefont{Knapp}},
  \bibinfo{author}{\bibfnamefont{O.}~\bibnamefont{G\"uhne}}, \bibnamefont{and}
  \bibinfo{author}{\bibfnamefont{H.~J.} \bibnamefont{Briegel}},
  \bibinfo{journal}{Phys. Rev. Lett.} \textbf{\bibinfo{volume}{99}},
  \bibinfo{pages}{250405} (\bibinfo{year}{2007}).

\bibitem[{\citenamefont{Vitagliano et~al.}(2011)\citenamefont{Vitagliano,
  Hyllus, Egusquiza, and T{\'o}th}}]{vitagliano2011spin}
\bibinfo{author}{\bibfnamefont{G.}~\bibnamefont{Vitagliano}},
  \bibinfo{author}{\bibfnamefont{P.}~\bibnamefont{Hyllus}},
  \bibinfo{author}{\bibfnamefont{I.~L.} \bibnamefont{Egusquiza}},
  \bibnamefont{and} \bibinfo{author}{\bibfnamefont{G.}~\bibnamefont{T{\'o}th}},
  \bibinfo{journal}{Phys. Rev. Lett.} \textbf{\bibinfo{volume}{107}},
  \bibinfo{pages}{240502} (\bibinfo{year}{2011}).

\bibitem[{\citenamefont{Troiani et~al.}(2013)\citenamefont{Troiani, Carretta,
  and Santini}}]{Troiani13}
\bibinfo{author}{\bibfnamefont{F.}~\bibnamefont{Troiani}},
  \bibinfo{author}{\bibfnamefont{S.}~\bibnamefont{Carretta}}, \bibnamefont{and}
  \bibinfo{author}{\bibfnamefont{P.}~\bibnamefont{Santini}},
  \bibinfo{journal}{Phys. Rev. B} \textbf{\bibinfo{volume}{88}},
  \bibinfo{pages}{195421} (\bibinfo{year}{2013}).

\bibitem[{\citenamefont{Dowling et~al.}(2004)\citenamefont{Dowling, Doherty,
  and Bartlett}}]{dowling+04pra}
\bibinfo{author}{\bibfnamefont{M.~R.} \bibnamefont{Dowling}},
  \bibinfo{author}{\bibfnamefont{A.~C.} \bibnamefont{Doherty}},
  \bibnamefont{and} \bibinfo{author}{\bibfnamefont{S.~D.}
  \bibnamefont{Bartlett}}, \bibinfo{journal}{Physical Review A}
  \textbf{\bibinfo{volume}{70}}, \bibinfo{pages}{062113}
  (\bibinfo{year}{2004}).

\bibitem[{\citenamefont{G{\"u}hne et~al.}(2005)\citenamefont{G{\"u}hne,
  T{\'o}th, and Briegel}}]{guhne+05NJOP}
\bibinfo{author}{\bibfnamefont{O.}~\bibnamefont{G{\"u}hne}},
  \bibinfo{author}{\bibfnamefont{G.}~\bibnamefont{T{\'o}th}}, \bibnamefont{and}
  \bibinfo{author}{\bibfnamefont{H.~J.} \bibnamefont{Briegel}},
  \bibinfo{journal}{New Journal of Physics} \textbf{\bibinfo{volume}{7}},
  \bibinfo{pages}{229} (\bibinfo{year}{2005}).

\bibitem[{\citenamefont{Troiani and Siloi}(2012)}]{troiani2012energy}
\bibinfo{author}{\bibfnamefont{F.}~\bibnamefont{Troiani}} \bibnamefont{and}
  \bibinfo{author}{\bibfnamefont{I.}~\bibnamefont{Siloi}},
  \bibinfo{journal}{Phys. Rev. A} \textbf{\bibinfo{volume}{86}},
  \bibinfo{pages}{032330} (\bibinfo{year}{2012}).

\bibitem[{\citenamefont{O'Connor and Wootters}(2001)}]{wooters2001entangled}
\bibinfo{author}{\bibfnamefont{K.~M.} \bibnamefont{O'Connor}} \bibnamefont{and}
  \bibinfo{author}{\bibfnamefont{W.~K.} \bibnamefont{Wootters}},
  \bibinfo{journal}{Phys. Rev. A} \textbf{\bibinfo{volume}{63}},
  \bibinfo{pages}{052302} (\bibinfo{year}{2001}).

\bibitem[{\citenamefont{Arnesen et~al.}(2001)\citenamefont{Arnesen, Bose, and
  Vedral}}]{arnsen+01prl}
\bibinfo{author}{\bibfnamefont{M.~C.} \bibnamefont{Arnesen}},
  \bibinfo{author}{\bibfnamefont{S.}~\bibnamefont{Bose}}, \bibnamefont{and}
  \bibinfo{author}{\bibfnamefont{V.}~\bibnamefont{Vedral}},
  \bibinfo{journal}{Phys. Rev. Lett.} \textbf{\bibinfo{volume}{87}},
  \bibinfo{pages}{017901} (\bibinfo{year}{2001}).

\bibitem[{\citenamefont{Gatteschi et~al.}(2007)\citenamefont{Gatteschi,
  Sessoli, and Villain}}]{Gatteschi}
\bibinfo{author}{\bibfnamefont{D.}~\bibnamefont{Gatteschi}},
  \bibinfo{author}{\bibfnamefont{R.}~\bibnamefont{Sessoli}}, \bibnamefont{and}
  \bibinfo{author}{\bibfnamefont{J.}~\bibnamefont{Villain}},
  \emph{\bibinfo{title}{Molecular nanomagnets}} (\bibinfo{publisher}{Oxford
  University Press}, \bibinfo{year}{2007}).

\bibitem[{\citenamefont{Affronte et~al.}(2007)\citenamefont{Affronte, Carretta,
  Timco, and P.}}]{Winpenny07}
\bibinfo{author}{\bibfnamefont{M.}~\bibnamefont{Affronte}},
  \bibinfo{author}{\bibfnamefont{S.}~\bibnamefont{Carretta}},
  \bibinfo{author}{\bibfnamefont{G.~A.} \bibnamefont{Timco}}, \bibnamefont{and}
  \bibinfo{author}{\bibfnamefont{W.~R.~E.} \bibnamefont{P.}},
  \bibinfo{journal}{Chem. Commun.} p. \bibinfo{pages}{1789}
  (\bibinfo{year}{2007}).

\bibitem[{\citenamefont{Lorusso et~al.}(2012)\citenamefont{Lorusso, Corradini,
  Ghirri, Biagi, del Pennino, Siloi, Troiani, Timco, Winpenny, and
  Affronte}}]{Lorusso12}
\bibinfo{author}{\bibfnamefont{G.}~\bibnamefont{Lorusso}},
  \bibinfo{author}{\bibfnamefont{V.}~\bibnamefont{Corradini}},
  \bibinfo{author}{\bibfnamefont{A.}~\bibnamefont{Ghirri}},
  \bibinfo{author}{\bibfnamefont{R.}~\bibnamefont{Biagi}},
  \bibinfo{author}{\bibfnamefont{U.}~\bibnamefont{del Pennino}},
  \bibinfo{author}{\bibfnamefont{I.}~\bibnamefont{Siloi}},
  \bibinfo{author}{\bibfnamefont{F.}~\bibnamefont{Troiani}},
  \bibinfo{author}{\bibfnamefont{G.}~\bibnamefont{Timco}},
  \bibinfo{author}{\bibfnamefont{R.~E.~P.} \bibnamefont{Winpenny}},
  \bibnamefont{and} \bibinfo{author}{\bibfnamefont{M.}~\bibnamefont{Affronte}},
  \bibinfo{journal}{Phys. Rev. B} \textbf{\bibinfo{volume}{86}},
  \bibinfo{pages}{184424} (\bibinfo{year}{2012}).

\bibitem[{\citenamefont{Siloi and Troiani}(2012)}]{siloi2012towards}
\bibinfo{author}{\bibfnamefont{I.}~\bibnamefont{Siloi}} \bibnamefont{and}
  \bibinfo{author}{\bibfnamefont{F.}~\bibnamefont{Troiani}},
  \bibinfo{journal}{Phys. Rev. B} \textbf{\bibinfo{volume}{86}},
  \bibinfo{pages}{224404} (\bibinfo{year}{2012}).

\bibitem[{\citenamefont{Candini et~al.}(2010)\citenamefont{Candini, Lorusso,
  Troiani, Ghirri, Carretta, Santini, Amoretti, Muryn, Tuna, Timco
  et~al.}}]{candini+10prl}
\bibinfo{author}{\bibfnamefont{A.}~\bibnamefont{Candini}},
  \bibinfo{author}{\bibfnamefont{G.}~\bibnamefont{Lorusso}},
  \bibinfo{author}{\bibfnamefont{F.}~\bibnamefont{Troiani}},
  \bibinfo{author}{\bibfnamefont{A.}~\bibnamefont{Ghirri}},
  \bibinfo{author}{\bibfnamefont{S.}~\bibnamefont{Carretta}},
  \bibinfo{author}{\bibfnamefont{P.}~\bibnamefont{Santini}},
  \bibinfo{author}{\bibfnamefont{G.}~\bibnamefont{Amoretti}},
  \bibinfo{author}{\bibfnamefont{C.}~\bibnamefont{Muryn}},
  \bibinfo{author}{\bibfnamefont{F.}~\bibnamefont{Tuna}},
  \bibinfo{author}{\bibfnamefont{G.}~\bibnamefont{Timco}},
  \bibnamefont{et~al.}, \bibinfo{journal}{Phys. Rev. Lett.}
  \textbf{\bibinfo{volume}{104}}, \bibinfo{pages}{037203}
  (\bibinfo{year}{2010}).

\bibitem[{\citenamefont{Auerbach}(1994)}]{auerbach}
\bibinfo{author}{\bibfnamefont{A.}~\bibnamefont{Auerbach}},
  \emph{\bibinfo{title}{Interacting electrons and quantum magnetism}}
  (\bibinfo{publisher}{Springer}, \bibinfo{year}{1994}).

\bibitem[{\citenamefont{Tsukerblat}(2006)}]{tsukerblat}
\bibinfo{author}{\bibfnamefont{B.}~\bibnamefont{Tsukerblat}},
  \emph{\bibinfo{title}{Group Theory in Chemistry and Spectroscopy}}
  (\bibinfo{publisher}{Dover Publication}, \bibinfo{address}{New York},
  \bibinfo{year}{2006}).

\bibitem[{\citenamefont{Caciuffo et~al.}(2005)\citenamefont{Caciuffo, Guidi,
  Amoretti, Carretta, Liviotti, Santini, Mondelli, Timco, Muryn, and
  Winpenny}}]{Caciuffo05}
\bibinfo{author}{\bibfnamefont{R.}~\bibnamefont{Caciuffo}},
  \bibinfo{author}{\bibfnamefont{T.}~\bibnamefont{Guidi}},
  \bibinfo{author}{\bibfnamefont{G.}~\bibnamefont{Amoretti}},
  \bibinfo{author}{\bibfnamefont{S.}~\bibnamefont{Carretta}},
  \bibinfo{author}{\bibfnamefont{E.}~\bibnamefont{Liviotti}},
  \bibinfo{author}{\bibfnamefont{P.}~\bibnamefont{Santini}},
  \bibinfo{author}{\bibfnamefont{C.}~\bibnamefont{Mondelli}},
  \bibinfo{author}{\bibfnamefont{G.}~\bibnamefont{Timco}},
  \bibinfo{author}{\bibfnamefont{C.~A.} \bibnamefont{Muryn}}, \bibnamefont{and}
  \bibinfo{author}{\bibfnamefont{R.~E.~P.} \bibnamefont{Winpenny}},
  \bibinfo{journal}{Phys. Rev. B} \textbf{\bibinfo{volume}{71}},
  \bibinfo{pages}{174407} (\bibinfo{year}{2005}).

\end{thebibliography}

\end{document}